\documentstyle[12pt]{article}
\begin{document}
\begin{center}
{\Large \bf On Quantum Nonlocality:\\
Using Prediction of a Distant Measurement Outcome} \\[1.5cm]
{\bf Vladimir S.~MASHKEVICH}\footnote {E-mail:
mash@gluk.apc.org}  \\[1.4cm]
{\it Institute of Physics, National academy
of sciences of Ukraine \\
252028 Kiev, Ukraine} \\[1.4cm]
\vskip 1cm

{\large \bf Abstract}
\end{center}

We assume that an event caused by a correlation between
outcomes of two causally separated measurements is, by
definition, a manifestation of quantum nonlocality, or
superluminal influence. An example of the Alice-Bob type
is given, with the characters replaced. The relationship
between quantum nonlocality and relativity theory is
touched upon.

\newpage

\section*{Introduction}

A recent paper by Stapp [1] has breathed new life into the
problem of quantum nonlocality, or superluminal influence.
As a result, a controversy has been aroused [2-11]. Opinions
differ widely: Quantum nonlocality exists and may be proved
using counterfactuals; quantum nonlocality exists but
the counterfactual proof is untenable; quantum nonlocality
does not exist.

Our opinion is the second one, so that an existence proof of
quantum nonlocality should be based on actual events. The aim
of the present paper is to propose such a proof. It goes without
saying that first and foremost a definition of quantum nonlocality
or, to be more precise, of its manifestation should be given.

We assume that an event caused by a correlation between outcomes
of two causally separated measurements is, by definition, a
manifestation of quantum nonlocality, or superluminal influence.

Given this definition, there is no difficulty in constructing an
existence proof. We choose that of the Alice-Bob type, replacing
the characters. Other examples, which relate to Bell inequalities,
are well known.

The relationship between quantum nonlocality and relativity
theory (for this problem see, e.g., [12]) is touched upon.

\section{Definition of quantum nonlocality via its
manifestation}

Nobody would deny that due to quantum entanglement there
exists a correlation between outcomes of two causally
separated measurements. But since the correlation does
not imply superluminal signals, not all treat it as quantum
nonlocality. In the long run, that is a matter of taste.
Be that as it may, it seems reasonable to define quantum
nonlocality via manifestations of the correlation.

By definition, we assume that an event caused by a correlation
between outcomes of two causally separated measurements is a
manifestation of quantum nonlocality, or superluminal
influence.

Here the term `event' has a standard relativistic meaning: An
event is localized in spacetime.

\section{Two civilizations}

There are two civilizations: aggressive ($A$) and intellectual
($I$). The time distance between them is
\begin{equation}
T\equiv T^{A}_{A-I-A}=T^{I}_{I-A-I}={\rm const}
\label{2.1}
\end{equation}
where $T^{A}$ stands for a time interval by $A$ clock and
$A-I-A$ for a light signal from $A$ to $I$ to $A$.

$A$ desires to destroy $I$. $A$ can send a destroying light
pulse with one of frequencies $\Omega_{i},\quad i=1,2,...,N,
\quad N\gg 1$. $I$ has $N$ mirrors, $M_{1},M_{2},...,M_{N}$.
The mirror $M_{i}$ reflects the pulse $\Omega_{i}$, so that
\begin{equation}
{\rm combination}\;(\Omega_{i},M_{i'})\; {\rm results \;
in\;destroying}\;\left\{
		  \begin{array}{rcl}
		   I\;{\rm for}\;i'\ne i\\
		   A\;{\rm for}\;i'=i. \\
		  \end{array}
		 \right.
\label{2.2}
\end{equation}
The setting-up time for a mirror is
\begin{equation}
T^{I}_{\rm setting}=\frac{1}{2}T-\tau^{I},\quad \tau^{I}\ll
\frac{1}{2}T.
\label{2.3}
\end{equation}

$A$ is corrupt to the last degree. $I$ has an excellent
secret service.

$A$ will send a pulse if $\Omega_{i}$ is unknown to $I$:
in view of $N\gg 1$, the risk is small.

\section{An order}

To get around the corruption and secret service, $A$ decides
that the choice of $\Omega_{i}$ should be a random event.
An order is given to a physical laboratory: At the time
$t^{A}_{\rm receiving}$, a quantum system ($A$ system) should
be received in a mixed state with a statistical operator
\begin{equation}
\rho^{A}=\frac{1}{N}\sum_{i}^{1,N}\left| Ai \right\rangle
\left\langle Ai \right|,\quad
\left\langle Ai \right|\left. Ai' \right\rangle=
\delta_{ii'},
\label{3.1}
\end{equation}
where
\begin{equation}
O^{A}\left| Ai \right\rangle=a_{i}\left| Ai \right\rangle.
\label{3.2}
\end{equation}
At the instant
\begin{equation}
t^{A}_{{\rm measuring}}=t^{A}_{{\rm receiving}}
\label{3.3}
\end{equation}
the observable $O^{A}$ will be measured with a result $a_{i}$,
and at the instant
\begin{equation}
t^{A}_{{\rm sending}}=t^{A}_{{\rm measuring}}
\label{3.4}
\end{equation}
a pulse $\Omega_{i}$ will be send.

\section{The order is fulfilled}

Due to an operation by $I$ secret service, the order is fulfilled
as follows. At the time $t^{A}_{{\rm receiving}}$ $A$ receives
$A$ system with $\rho^{A}$ given by eq.(\ref{3.1}), where
\begin{equation}
\rho^{A}={\rm Tr}_{I}\rho^{AI},
\label{4.1}
\end{equation}
\begin{equation}
\rho^{AI}=\left| AI \right\rangle\left\langle AI \right|,\quad
\left| AI \right\rangle=\frac{1}{\sqrt{N}}\sum_{i}^{1,N}
\left| Ai \right\rangle\otimes \left| Ii \right\rangle,
\quad \left\langle Ii \right|\left. Ii' \right\rangle=
\delta_{ii'},
\label{4.2}
\end{equation}
\begin{equation}
O^{I}\left| Ii \right\rangle=b_{i}\left| Ii \right\rangle.
\label{4.3}
\end{equation}
$I$ receives $I$ system at the time $t^{I}_{{\rm receiving}}$
such that
\begin{equation}
t^{I}_{{\rm coming}}=t^{I}_{{\rm receiving}}+\frac{1}{2}T
\label{4.4}
\end{equation}
where $t^{I}_{{\rm coming}}$ stands for the instant
of the pulse coming.

\section{The result}

The observable $O^{I}$ is measured at the instant
\begin{equation}
t^{I}_{{\rm measuring}}=t^{I}_{{\rm receiving}}
\label{5.1}
\end{equation}
with a result $b_{i}$ corresponding to $a_{i}$. The mirror
$M_{i}$ is set up by the time
\begin{equation}
t^{I}_{{\rm receiving}}+T^{I}_{{\rm setting}}=
t^{I}_{{\rm receiving}}+\frac{1}{2}T-\tau^{I}<
t^{I}_{{\rm coming}}.
\label{5.2}
\end{equation}
The aggressor $A$ is destroyed.

The event referred to in Sec. 1 is the impact of the pulse
$\Omega_{i}$ on the mirror $M_{i}$.

We may say that $I$ has used the prediction that the outcome
of the distant measurement of $O^{A}$ is $a_{i}$.

\section{Quantum nonlocality and relativity theory}

We would not say that quantum nonlocality contradicts special
relativity: the situation is not so simple. Quantum nonlocality
implies an additional structure of spacetime, which is absent
in special relativity. The structure is this: The hypersurface
of a quantum jump is that of a constant value of cosmic
time [13].

\section*{Acknowledgement}

I would like to thank Stefan V. Mashkevich for helpful
discussion.

\end{document}